\documentclass[sigplan,10pt,screen]{acmart}
\acmConference[PL'18]{ACM SIGPLAN Conference on Programming Languages}{January 01--03, 2018}{New York, NY, USA}
\acmYear{2018}
\acmISBN{} 
\acmDOI{} 
\startPage{1}
\setcopyright{none}
\usepackage{paper}

\newacronym{AI}{AI}{abstract interpretation}
\newacronym{CFG}{CFG}{control flow graph}
\newacronym{DFA}{DFA}{data-flow analysis}
\newacronym{DFS}{DFS}{depth-first search}
\newacronym[longplural={difference bounded matrices}]{DBM}{DBM}{difference bounded matrix}
\newacronym{JVM}{JVM}{Java Virtual Machine}
\newacronym{LIA}{LIA}{linear integer arithmetic}
\newacronym{NIA}{NIA}{non-linear integer arithmetic}
\newacronym{TVPI}{TVPI}{two variables per inequality}
\newacronym{uTVPI}{uTVPI}{unit two variables per inequality}
\newacronym{SMT}{SMT}{satisfiability modulo theories}
\newacronym{SMT-LIB}{SMT-LIB}{satisfiability modulo theories library}
\newacronym{StInG}{StInG}{Stanford Invariant Generator}

\newacronym{FC}{FC}{Full Comparison}
\newacronym{FS}{FS}{full state}
\newacronym{FG}{FG}{Full Graph}
\newacronym{CC}{CC}{Connect Components}
\newacronym{NN}{NN}{Node Neighbors}
\newacronym{MN}{MN}{Minimal Neighbors}

\begin{document}


\title{Minimal Comparison of Octagonal Abstract Domains}

\author{Kenny Ballou}
\orcid{0000-0002-6032-474X} 
\affiliation{
  \department{Computer Science and Engineering}
  \institution{California State University San Marcos}
  \city{San Marcos}
  \state{California}
  \country{United States}
}
\email{kballou@csusm.edu}

\author{Elena Sherman}
\orcid{0000-0003-4522-9725} 
\affiliation{
  \department{Computer Science}
  \institution{Boise State University}
  \city{Boise}
  \state{Idaho}
  \country{United States}
}
\email{elenasherma@boisestate.edu}

\begin{abstract}

  Numerical abstract domains vary in their expressiveness; more expressive
  domains like Zones yield more precise invariants than Intervals.  A
  comprehensive approach to selecting abstract domains is a minimal comparison
  of abstract states.  However, to be effective, it requires abstract states to
  be free of spurious constraints.  While previous work developed spurious
  constraint elimination for Zones, this work introduces a novel algorithm for
  eliminating such constraints for Octagons.

  We evaluate our approach by comparing the precision of 6,930 invariants from
  different abstract domains.  Our results show that the minimal comparison
  reclassifies many invariants as equivalent, thus reducing the impact of
  Octagons' expressiveness on invariant precision.

\end{abstract}



\maketitle{}

\section{Introduction}\label{sec:intro}%

The tradeoff between expressiveness and efficiency of different numerical
abstract domains like Zones~\cite{mine-2001-new-numer},
Octagons~\cite{mine-2006-octag-abstr-domain} or Intervals, requires an
understanding of the impact of precision loss when choosing a more efficient,
but less expressive abstract domain.  While related
work~\cite{apel-2013-domain-types, bodden-2018-self-adapt, howe-2009-logah,
  logozzo-2010-pentag} establish relations between program structural
characteristics and an appropriate abstract domain choice, this information
alone is insufficient to comprehensively compare domains.

Traditionally comparing two abstract domains, \(D_1\) and \(D_2\), where the
former is more precise than the latter, involves comparing their invariants
\(I_1\) and \(I_2\) at the same program locations.  If the invariants of
\(D_1\) dominate in precision then one concludes a high impact on precision
gain if \(D_1\) is selected.  Other works~\cite{casso-2020-comput-abstr,
  giacobazzi-2023-how-fittin, schwarz-2023-clust-relat}, create a distance
metric to compare states.  However, in the context of \gls{DFA}, previous
work~\cite{ballou-2023-ident-minim, ballou-2023-minim-compar} show that a novel
minimal comparison technique gives more insights into the precision gain or
loss when choosing \(D_1\) or \(D_2\), respectively.

The minimal comparison leverages the fact that at each computational step,
\gls{DFA} applies incremental changes to an abstract state, i.e., a transfer
function only modifies a part of an abstract state.  While other techniques use
incrementality to improve performance~\cite{ballou-2022-increm-trans,
  gange-2021-fresh-look, jourdan-2017-spars-preser-algor-octag}, the line of
work on minimal comparison uses incremental updates to identify the changed
portion of the abstract state and hence, compare only the updated parts of the
resulting invariants.  This eliminates cases when an improvement in precision
in one statement propagates through the entire computation.  For example, in
this sequence of statements 1:\texttt{x = y + 2}, 2:\texttt{z = 2}, 3:\texttt{w
  = 4 - z} using Zones increases precision after statement 1: but for the other 
two statements, Zones computes the same invariants as Intervals for the \(z\)
and \(w\) part of abstract states.  Yet, comparison of the entire states
results in Zones being globally more precise.

To effectively apply the minimal change
algorithm~\cite{ballou-2023-ident-minim}, it is imperative to remove spurious
dependencies between variables in a fully closed form of an abstract state.
While previous work developed such an algorithm for
Zones~\cite{ballou-2023-ident-minim, ballou-2023-minim-compar}, it cannot,
however, be applied to Octagons because of the substantial differences in their
state representations.  Thus, in this work we present a novel algorithm for
removing spurious dependencies for fully closed Octagon abstract states, which
we describe in Section~\ref{sec:approach}.  Equipped with this new algorithm,
we, in Section~\ref{sec:results}, perform extensive empirical evaluation on how
comparing minimal changes of invariants impacts comparison of abstract domains
within the context of \gls{DFA}.  We discuss related works in
Section~\ref{sec:related} and finally conclude with future works in
Section~\ref{sec:concl}.


\section{Background}\label{sec:backgrd}%

\subsection{Zones}

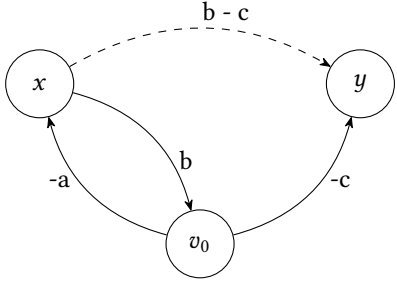
\begin{figure}[ht]
  \begin{tikzpicture}[%
  auto,
  node distance=3.0cm,
  var/.style={draw, circle, minimum size=0.9cm},
  dvar/.style={double},
  bound/.style={draw, -{Stealth[round]}},
  inferred/.style={dashed}]

  \node[var] (v0) {\(v_0\)};
  \node[var, above left of=v0] (x) {\(x\)};
  \node[var, above right of=v0] (y) {\(y\)};

  \draw[bound] (v0) edge [bend left] node[left, pos=0.6] {-a} (x);
  \draw[bound] (x) edge [bend left] node[right, pos=0.7] {b} (v0);
  \draw[bound] (v0) edge [bend right] node[right, pos=0.6] {-c} (y);

  \draw[bound, dashed] (x) edge [bend left] node[above, pos=0.6] {b - c} (y);
\end{tikzpicture}

  \caption{Directed weighted graph representation of unit-difference
    constraints of a Zone abstract state.}%
  \label{fig:zones}
\end{figure}

The Zones domain~\cite{mine-2001-new-numer} represent predicates via a strict
unit difference formula, extending the intervals domains with the form
\(x - y \le c\), where \(x\) and \(y\) are program variables, and \(c\) is a
numerical constant from one of \(\{\mathbb{Z}, \mathbb{Q}, \mathbb{R}\}\).  Interval valued constraints
are encoded by adding a special variable, typically denoted \(v_0\), such that
its value is always zero.  Thus,
\(x - v_0 \le b \implies x - 0 \le b \implies x \le b\) and
\(v_0 - x \le -a \implies 0 - x \le -a \implies x \ge a\) are sufficient to encode
the interval for \(x\): \(x \mapsto [a, b]\).  Figure~\ref{fig:zones} depicts the
directed weighted graph representation of a Zone domain for the interval
constraints \(x \mapsto to [a, b] \land y \mapsto to \cointv{c}{+\infty}\).

\subsection{Octagons}

The Octagons domain extends the unit difference of
Zones~\cite{mine-2001-new-numer} by encoding unit constraints over program
variables in the forms \(\pm x \pm y \le c\) and \(\pm x \le 2c\).  The symbols
\(x, y\) represent program variables, and \(c\) is a numerical
constant\footnote{Our work focuses on integer constraints, but should easily
  extend to \(\mathbb{Q}\) and \(\mathbb{R}\) since integers require additional care the others
  do not.}.  To efficiently manipulate such constraints, researchers proposed
an encoding~\cite{chawdhary-2018-increm-closin-octag, gange-2021-fresh-look,
  mine-2006-octag-abstr-domain} where variables are translated to a pair of
constraint nodes, denoted here as \(x^{\pm}\) and \(y^{\pm}\), which can be
expressed as a directed, weighted graph.  The nodes of the graph represent the
constraint variables, and the edge weight denotes the difference constraint
between the variables.  The direction of the edge represents the ordering of
the formula.  Moreover, the graphical representation has an isomorphic
representation as a \gls{DBM}, denoted throughout as \(M\).  The standard
Octagon encoding rules are shown in Figure~\ref{fig:octagon-rules}.

\begin{figure}[t]
  \include{images/octagon-encoding-rules}
  \caption{Standard encoding rules for converting program variable space
    predicates and converting them to octagonal variable space.}%
  \label{fig:octagon-rules}
\end{figure}

For example, consider the code snippet: \texttt{if (y >= a)\{ x = b; \ldots\}},
where \texttt{a, b} are constants and \texttt{x, y} are program variables.  The
graph and \gls{DBM} in Figure~\ref{fig:octagon-example} shows the octagonal
encoding of the state on the true branch, i.e.,
\(y \ge a \leadsto y^- - y^+ \le -2a\) and
\(x = b \leadsto x^+ - x^- \le 2b \land x^- - x^+ \le -2b\).  After processing \texttt{x =
  b}, \gls{DFA} modifies the state in Figure~\ref{fig:octagon-graph-1}, with
the newly added constraints depicted in the highlighted box.  We collect these
newly added edges into a set, \(de\), which represents the changed edges as it
relates to the abstract state.  In this case, the set \(de\) is equal to the
following \(\{(x^+, x^-, 2b), (x^-, x^+, -2b)\}\), which is equivalent to the
logical expression, \(x^+ - x^- \le 2b \land x^- - x^+ \le -2b\).

\begin{figure}[t]
  \centering
  \begin{subfigure}[b]{0.49\textwidth}
    \centering{}
    \begin{tikzpicture}[%
  auto,
  node distance=3.0cm,
  var/.style={draw, circle, minimum size=0.9cm},
  dvar/.style={double},
  bound/.style={draw, -{Stealth[round]}},
  infEdge/.style={dashed}]

  \draw[fill=gray!10, rounded corners] (-0.8, -3.8) rectangle (0.8, 0.8);
  \node[var, dvar] (i) {\({x^+_{i = 0}}\)};
  \node[var, dvar, below of=i] (ibar) {\({x^-_{\bar{\imath} = 1}}\)};
  \node[var, right of=i] (j) {\({y^+_{j = 2}}\)};
  \node[var, below of=j] (jbar) {\({y^-_{\bar{\jmath} = 3}}\)};

  \draw[bound] (i) edge[bend left] node[left, pos=0.66] {2b} (ibar);
  \draw[bound] (ibar) edge[bend left] node[right, pos=0.66] {-2b} (i);
  \draw[bound] (jbar) edge[] node[pos=0.66] {-2a} (j);

  \draw[bound, infEdge] (i) edge[above] node[] {b - a} (j);
  \draw[bound, infEdge] (jbar) edge[below] node[] {b - a} (ibar);

  \draw[bound, infEdge] (ibar) edge[] node[sloped, above, pos=0.66] {-a - b} (j);
  \draw[bound, infEdge] (jbar) edge[] node[sloped, below, pos=0.25] {-a - b} (i);
\end{tikzpicture}
    \caption{Fully closed graphical representation of an octagonal state.}%
    \label{fig:octagon-graph-1}
  \end{subfigure}
  \hfill
  \begin{subfigure}[b]{0.49\textwidth}
    \centering{}
    \begin{tikzpicture}[
  inferred/.style={dashed, rounded corners},
  newcon/.style={opacity=0.3, fill=gray!20, rounded corners},
  row labels/.style={
    matrix of math nodes,
    nodes in empty cells,
    nodes={minimum size=5mm, anchor=center}
  },
  col labels/.style={matrix of math nodes,
    nodes in empty cells,
    nodes={minimum size=10mm, anchor=center}
  },
  dbm/.style={matrix of math nodes,
    nodes in empty cells,
    nodes={minimum size=5mm, anchor=center},
    left delimiter=[,
    right delimiter=]
  }]
  \matrix[dbm] (dbm) {
    0       & 2b    & b - a  & +\infty \\
    2b      & 0     & -a - b & +\infty \\
    +\infty      & +\infty    & 0      & +\infty \\
    -a - b  & b - a & -2a    & 0  \\
  };

  \matrix[col labels, above=-0.4cm of dbm] {
    {x}^+_{i=0} & {x}^-_{\bar{\imath}=1} & {y}^+_{j=2} & {y}^-_{\bar{\jmath}=3}\\
  };

  \matrix[row labels, left=0.25cm of dbm] {
    {x}^+ \\
    {x}^- \\
    {y}^+ \\
    {y}^- \\
  };

  \draw[newcon] (dbm-1-2.north west) rectangle (dbm-1-2.south east);
  \draw[newcon] (dbm-2-1.north west) rectangle (dbm-2-1.south east);
  \draw[inferred] (dbm-1-3.north west)[xshift=-1mm] rectangle (dbm-2-3.south east);
  \draw[inferred] (dbm-4-1.north west) rectangle (dbm-4-2.south east);
\end{tikzpicture}

    \caption{Isomorphic \acrfull{DBM} representation of the same octagonal
      state.}%
    \label{fig:octagon-dbm-1}
  \end{subfigure}
  \caption{Fully, strongly closed Octagons represented as a weighted, directed
    graphs and its equivalent representation as a \acrfull{DBM}.  The state
    contains \(y \ge a \land x = b\) as well as inferred constraints
    \(x - y \le b + a \land -x -y \le a - b\).  The double circles denote changed
    variables, dashed edges denote inferred edges through closure.}%
  \label{fig:octagon-example}
  \vspace{-1.5em}
\end{figure}

\subsection{Symbolic Predicates Abstract Domain}

For evaluating our comparison, we use a so-called symbolic predicate domain to
represent an instance of an incomparable domain.  The Symbolic Predicates
abstract domain extends the well-known Predicates domain with arbitrary
inequality formula.  These inequalities are automatically inferred through the
program's transfer functions when performing Predicate
analysis~\cite{sherman-2015-exploit-domain}.  Furthermore, since these
inequalities are arbitrary, they can add exact, relational information to the
predetermined set of predicates.  For example, using the Sign domain,
interpreting the expression \(y = -2x\), with an incoming state of
\(x \to \{-\} \land y \to \top\), becomes \(x \to \{-\} \land y \to \{+\}\).  Augmenting
with symbolic information, the expression after computing its transfer function
becomes \(x \to \{-\} \land y \to \{+\} \land y = -2x\), a precise relational
representation of the assignment.  We simply use this domain as an
instantiation of an incomparable domain to Octagons.

\subsection{Additional Notations}

While we use some common notation throughout this work, we define some of the
notation for clarity.  When using the matrix notation for \glspl{DBM}, we often
use index variables \(i\), \(j\), \(\bar{\imath}\), \(\bar{\jmath}\).  Previous work
has used the latter two, to represent the ``negative'' dual of the
former~\cite{chawdhary-2018-increm-closin-octag}.  That is, we denote variable
pairs within the octagonal encoding using \(i\) and \(\bar{\imath}\) for each real
program variable such that \(i \oplus \bar{\imath} \equiv 1\), where \(\oplus\) is the exclusive or
operator.  Furthermore, we can recover any index's pair by flipping operands:
\(i \oplus 1 \equiv \bar{\imath}\).  Thus, an edge between \(i\) and
\(\bar{\imath}\) represents the interval edges between a variable within an Octagon
constraint system.  Moreover, these notations are equivalent to the even/odd
\(2i_x\) and \(2i_x + 1\), encoding described by Min\'{e}
et~al.~\cite{mine-2006-octag-abstr-domain}.

\subsection{Spurious Constraints}

The critical step of finding a minimal state change is to precisely identify
all variables, \(\Delta{}vars\), affected by the newly added constraint \(de\).  To
do so, previous work by \citeauthor{ballou-2023-ident-minim} explores the
neighborhood of the changed variables \(dv\) of \(de\) within the
graph~\cite{ballou-2023-ident-minim}.  In our example, the gray box includes
all affected variables \(\Delta{}vars\), which in this case is only the changed
variables \(dv\) since there are no edges connecting them to \(y^+\) and
\(y^-\).

However, when an abstract state is in a canonical, fully closed form, those
variables become connected through implicit constraints shown as dashed edges
in Figure~\ref{fig:octagon-graph-1}, hence marking the variables \(y^{\pm}\) as
changed too.  We call these edges \emph{spurious} since they do not further
constrain the bounded region, but also decrease the precision of finding a
minimal state change.  Since the fully closed form is imperative for running
\gls{DFA}, e.g., for comparing states, we propose a new algorithm for removing
such spurious constraints in Octagon abstract states.


\section{Minimal Comparison}\label{sec:approach}%

\begin{figure}[t]
  \centering
  \begin{algorithm}[H]
      \caption{Identify minimally changed variables within an Octagon.}%
      \label{alg:min-delta-vars}
      \input{algorithms/min-delta-vars}
  \end{algorithm}
\end{figure}

Algorithm~\ref{alg:min-delta-vars} extends the algorithm from previous
work~\cite{ballou-2023-ident-minim} to identify minimally changed variables
\(\Delta{}\text{vars}\) when an abstract state \(M'\) is updated to a new state
\(M\) after processing a set of constraints \(de\), where each constraint
contains a triple of variables \((i, j, w)\).  After \(\Delta{}\text{vars}\) are
identified, then only constraints containing those variables are selected for
comparison.  Thus, the smaller the set of changed variables that
Algorithm~\ref{alg:min-delta-vars} returns the more accurate the comparison.

The algorithm takes the updated fully closed state encoded as a \gls{DBM} \(M\)
and the constraint set \(de\) associated with the update, where \(de\) is the
set of incoming, modified constraints.  First, on
lines~\ref{alg:min-delta-vars-init-begin}--\ref{alg:min-delta-vars-init-end},
the algorithm iterates over each constraint to determine whether \(i\) or \(j\)
indeed changed.  Next, on
lines~\ref{alg:min-delta-vars-select-begin}--\ref{alg:min-delta-vars-select-end},
the algorithm finds variables that are affected by the directly updated
variables \(dv\).  Since \(M\) is fully closed, it requires only one
``forward'' step and one ``backward'' step in the graph's representation of
\(M\).  Here \(N\) denotes the number of \emph{program variables} and
\(|M| = 2N \times 2N\), as per standard Octagon encoding that doubles the variable
space~\cite{mine-2006-octag-abstr-domain}.

However, when applied to fully closed states,
Algorithm~\ref{alg:min-delta-vars} is ineffective in identifying minimal
changes because such states contain many spurious constraints for variables
that encode interval values, similar to \(x\) and \(y\) from our example in
Figure~\ref{fig:octagon-example}.  That is, as part of the closure operation
and necessary operation, many variables become spuriously connected.  In our
example, \(x\) becomes spuriously coupled with \(y\) in the constraint
\(x - y \le b + a\).  Algorithm~\ref{alg:min-delta-vars} identifies \(x\) and
\(y\) as part of the \(\Delta{}\text{vars}\) set and uses both for comparison,
instead of just \(x\).  Therefore, we need a technique which removes these
\emph{spurious} constraints in \(M\) before passing this \gls{DBM} to
Algorithm~\ref{alg:min-delta-vars}.

\begin{figure}[t]
  \centering
  \begin{algorithm}[H]
    \input{algorithms/w0z-octag}
    \caption{Constraint reduction algorithm for Octagons}%
    \label{alg:w0z-octag}
  \end{algorithm}
\end{figure}

We define a \emph{spurious constraint} as a redundant constraint coupling
variables within a \gls{DBM} \(M\).  A positive-negative pair of variables
within the octagonal \gls{DBM} represent the interval bounds of a program
variable, e.g., \((x^{+}_{i}, x^{-}_{\bar{\imath}}, 2b) \equiv x \le b\), where
\(i \oplus \bar{\imath} \equiv 1\); dually for the other direction.  Due to the closure
operations, pairs of variables become coupled in a constraint.  However, if a
constraint is a result of these interval constraint closures, then we label it
as \emph{spurious}.  Formally, a constraint \(M_{i,j}\) in an Octagon matrix
\(M\) is spurious when the following two conditions hold:

\begin{enumerate}
\item{\(i \neq j \land i \oplus j \cancel{\equiv} 1\), and}
\item{\(M_{i,j} \ge {\left(M_{i,\bar{\imath}} + M_{\bar{\jmath},j}\right)}/2\).}
\end{enumerate}

The first condition ensures that it is not itself an interval constraint.  For
example, we do not consider constraints such as \(M_{i, \bar{\imath}}\).  The second
condition checks that a constraint over two variables is less or equally
constraining than its strong closure~\cite{mine-2006-octag-abstr-domain}.  If
this constraint is stronger, it is more restrictive than the interval values
alone.

Our novel Algorithm~\ref{alg:w0z-octag} removes spurious constraints for fully
closed Octagons states, i.e., it removes all recoverable constraints.  To do
so, the algorithm iterates over all indices of the closed \gls{DBM} \(M\),
lines~\ref{alg:w0z-begin}--\ref{alg:w0z-begin-2}, skipping interval valued
constraints, line~\ref{alg:w0z-skip-intv}.  Here, as before, \(N\) is the
number of \emph{program variables}.  In
lines~\ref{alg:w0z-spurious-beg}--\ref{alg:w0z-spurious-end}, if the constraint
is \emph{spurious}, the algorithm sets its value to \(+\infty\).

The runtime for Algorithm~\ref{alg:w0z-octag} is \(O(N^2)\).  Since the
\texttt{for} loops are definite with increasing indices, assuredly the
algorithm terminates.  Furthermore, the correctness of the algorithm is given
by its post-condition where all removed constraints are recoverable by
reapplying (strong) closure.


\section{Experiments and Results}\label{sec:results}%

To evaluate the effectiveness of the spurious constraint reduction algorithm
for Octagons, we compared the precision of invariants, via logical implication,
computed by \gls{DFA} using Octagons against those from other abstract domains.
To establish the baseline data, we compare entire invariants, i.e., full state
comparison or \gls{FC}.  Next, we minimally compare invariants using \gls{MN}
by selecting only the constraints of changed variables by leveraging
Algorithm~\ref{alg:min-delta-vars}.  Lastly we apply \gls{MN} on invariants
that underwent the spurious constraint reduction as described in
Algorithm~\ref{alg:w0z-octag} (MN + Reduction).

We divide experiments into comparisons of Octagons with Zones and Intervals,
i.e, comparable numerical domains and an incomparable Symbolic Predicates
domain.  The Symbolic Predicates domain~\cite{sherman-2015-exploit-domain} used
in this study includes the following initial disjoint predicate elements:
\(\{\ocintv{-\infty}{-5}\), \(\ocintv{-5}{-2}\), \(-1\), \(0\), \(1\),
\(\cointv{2}{5}\), \(\cointv{5}{+\infty}\}\).  This predicate domain was derived
from values identified from a previous empirical
study~\cite{collberg-2007-empir-study}.

While specialized predicates for each program would likely improve precision
for the symbolic Predicates domain, we chose a generic domain since our
interest lies in the trends within the logical entailment results between
domain instances.  Moreover, since our comparison study includes the
\emph{symbolic} component derived by the analyzer, the Predicates domain
already includes highly precise information.

Each experiment uses subject programs consisting of \(192\) Java methods from
previous research~\cite{ballou-2026-java-artif, sherman-2015-exploit-domain},
each exhibiting various levels of complexity.  The total number of invariants
is \(6,930\) compared in eight different configurations.  We used a
supercomputer and an existing \gls{DFA}
framework~\cite{ballou-2025-static-iced-tea, sherman-2018-redes-soots} to
compute all analyses.

\subsection{Comparable Domains}

Table~\ref{tab:octagons} shows the results of comparing Octagons (\(O\))
against Zones (\(Z\)) and Intervals (\(I\)).  Since the domains are ordered, an
invariant of \(I\) or \(Z\) can only be as precise (\(\equiv O\)) or less precise
than Octagons (\(\prec O\)).  The table includes the counts of each comparison
classification for each comparison technique.

\begin{table}[t]
  \caption{Intervals and Zones vs Octagons invariants comparison}%
  \label{tab:octagons}
  \centering
  \begin{tabular}{l@{\hspace{1.5em}}cc@{\hspace{2em}}cc}
  \toprule
  \textbf{Technique} & \(I \equiv O\) & \(I \prec O\) & \(Z \equiv O\) & \(Z \prec O\) \\[0pt]
  \midrule
  Full Comparison & 1523 & 5407 & 5138 & 1792 \\[0pt]
  \midrule
  MN & 3894 & 3036 & 6183 & 747 \\[0pt]
  \midrule
  MN + Reduction & 4439 & 2491 & 6527 & 403 \\[0pt]
  \bottomrule
\end{tabular}

\end{table}

The data shows that using full comparison, \(86\%\) of Octagon invariants are
more precise than Interval invariants.  However, using only minimization, the
proportion reduces to \(44\%\).  In MN+Reduction this proportion drops to
\(36\%\).  A similar trend occurs when comparing octagonal invariants to zonal
invariants: \(26\%\) to \(11\%\) to \(6\%\), respectively.

\subsection{Incomparable Domains}

Table~\ref{tab:octagons-predicates} shows the comparison results between
Symbolic Predicate and Octagon invariants.  Since these domains are
incomparable, the table has three additional classifications: \(O \succ P\),
Octagons less precise than Symbolic Predicates, and \(O \prec \succ P\), where both
invariants are incomparable.  A third column, \(O~?~P\), represents instances
where our solver, Z3~\cite{moura-2008-z3}, returned \texttt{UNKNOWN}\@.  For
this experiment, we used the iterative algorithm from previous
work~\cite{ballou-2023-minim-compar} for comparing incomparable domains and
omit data for \(MN\) experiment.

\begin{table}[t]
  \caption{Octagons and Symbolic Predicates}%
  \label{tab:octagons-predicates}
  \centering
  \begin{tabular}{lccccc}
  \toprule
  \bf Comparison & \(O \equiv P\) & \(O \prec P\) & \(O \succ P\) & \(O \prec\succ P\) & \bf \(O~?~P\) \\[0pt]
  \midrule
  Full & 1179 & 3082 & 499 & 2159 & 11 \\[0pt]
  \midrule
  MN + Reduc & 3768 & 2514 & 280 & 368 & 0 \\[0pt]
  \bottomrule
\end{tabular}

\end{table}

The data reveals that the number of equivalent invariants went up from \(17\%\)
to \(54\%\).  The trends from other classifications are decreasing: from
\(44\%\) to \(36\%\) for Symbolic Predicates more precise, from \(7\%\) to
\(4\%\) for Octagons more precise, from \(31\%\) to simply \(5\%\) for
incomparable.  Critically, minimal comparison resolved all \(11\) queries that
previously returned \texttt{UNKNOWN}, a direct result of the reduced query
size.

\subsection{Discussion}

The data from the tables suggests that the spurious reduction technique allows
for more comprehensive comparison between abstract domains.  That is, our
reduced and minimal technique of comparison of invariants demonstrates a finer
granularity of precision performance, especially for Symbolic Predicates.
Evaluations show that the impact of Octagons on invariant precision might not
be that significant.  These findings substantiate arguments from previous work
about weakly-relational domains computing predominately interval
invariants~\cite{gange-2021-fresh-look, gange-2016-exploit-spars}.



\section{Related Work}\label{sec:related}%

Research presenting new techniques or new abstract domains tend to use one of
two methods for comparison: either logical
entailment~\cite{mine-2004-weakl-relat, sherman-2015-exploit-domain}, or
property or invariant capture~\cite{gange-2021-fresh-look,
  gurfinkel-2010-boxes, howe-2009-logah, laviron-2008-subpol,
  logozzo-2010-pentag}.  Our work fits into the former group of work.  However,
this work extends previous work~\cite{ballou-2023-ident-minim,
  ballou-2023-minim-compar} work by providing algorithms for identifying
minimal subsets of Octagons.

In general, we notice an increasing amount of research in measuring and
quantifying the differences between various abstract
domains~\cite{ballou-2023-ident-minim, ballou-2023-minim-compar,
  casso-2020-comput-abstr, giacobazzi-2002-domain-compr,
  giacobazzi-2023-how-fittin, schwarz-2023-clust-relat}.  While our work does
not make any substantive arguments about distance between Intervals, Zones, and
Octagons, we do present several techniques which we believe complement the
current research.



\section{Conclusion}\label{sec:concl}%

In this work, we presented a reduction algorithm for Octagons that leverages
the incremental nature of \gls{DFA}.  Combining with previous work on minimally
comparing abstract domains, the reduction provides a more granular view of the
precision benefits of Octagons compared to other numerical domains.  Using
real-world programs, we empirically evaluated the invariant precision of
Octagons to Zones, Intervals, and Symbolic Predicates.  The results show that a
significant majority of invariants computed by Octagons are interval-valued,
experimentally confirming anecdotal comments of other previous
work~\cite{gange-2021-fresh-look}.

As noted previously, the reduction algorithm combined with work from Larsen
et~al.~\cite{larsen-1997-effic-verif, larsen-2003-compac-data} would complement
each other well in a model-checking context.  Furthermore, an interesting
future direction of this work would be exploring the interaction of the
techniques for identifying minimal changes within weakly-relational domains
complements research attempting to create metrics around domain precision.
Finally, it remains an open question whether the set of minimal changes can be
used to improve the efficiency of \gls{DFA}, by reducing the number of
comparisons needed to determine subsumption or equality between Octagon states
during the fixed point computation.




\nocite{tange-2024-gnu-paral}
\bibliography{bibfile, bibfile01}

\end{document}